\documentstyle[12pt]{article}
\input epsf.sty
\newcommand{\ba}{\begin{eqnarray}}
\newcommand{\ea}{\end{eqnarray}}
\newcommand{\be}{\begin{equation}}
\newcommand{\ee}{\end{equation}}

\begin{document}

\title{\bf
Nuclear Production of Charmonium \\ at ELFE Energies
}

\vspace{1 cm}
\author{
S. Gardner\\
\mbox{}\\
Department of Physics and Astronomy\\
University of Kentucky\\
Lexington, KY 40506-0055, USA
}
\date{}
\maketitle

\vspace{10 cm}
\noindent
December 1995\\
UK 95-15 \\
nucl-th/9512006

\vspace{3 cm}
\noindent
Invited talk presented at the ELFE Summer School on Confinement Physics,
July 22-28, 1995, Cambridge, England.

\newpage
\setcounter{page}{1}

\mbox{}
\vspace{3 cm}
\begin{center}
{\Large{\bf
NUCLEAR PRODUCTION OF CHARMONIUM \\
AT ELFE ENERGIES}}\\
\vspace{1.5 cm}
{\large S. Gardner\footnote{Present Address:
Department of Physics and Astronomy,
University of Kentucky,
Lexington, KY 40506-0055, USA.}
}\\
\medskip
{Nuclear Theory Center\\
Indiana University Cyclotron Facility\\
2401 Milo B. Sampson Lane\\
Bloomington, IN 47405, USA }
\\
\vspace{3.0 cm}
\large{\bf ABSTRACT}
\end{center}

I discuss the diffractive production of charmonium from nuclei
and the anomalous nuclear attenuation --- termed
``color transparency'' --- attributed to such processes,
focusing on the few tens of GeV energy regime germane
to the ELFE project.
A critical review of
color transparency calculations is presented, and the major sources
of uncertainty are emphasized.

\vspace{1.0 cm}

\section{Introduction --- why charmonium?}

In diffractive charmonium production, the produced
charmonium states emerge with
nearly all the momentum kinematically accessible to them. In
a semi-inclusive
photoproduction experiment, in which only charmonium is
detected, the constraint of diffractive production requires that
the transverse
component of the charmonium momentum be small.
Diffractive production from a nuclear target is of interest, as in
this kinematic regime the nuclear attenuation can vary markedly from
standard estimates~\cite{brodmuel88}.
That is,
the longitudinal momentum transfer to convert
one charmonium state to another under interactions with the nuclear medium
vanishes as the charmonium momentum
increases~\cite{goodwalk}, so that the
amplitudes
of the individual charmonium states ({\it e.g.}, $J/\psi, \psi'$)
are scrambled in their exit through the nucleus. In this manner
the nuclear attenuation can suffer large excursions from that of a
Glauber model estimate, in which the charmonium's
internal structure is neglected.
Such anomalous nuclear attenuation effects have been discussed
in the context of processes such as the
$(e,e'p)$ reaction at large momentum transfer and
small missing energy and termed ``color transparency''~\cite{ctbrodmuel},
as produced
states of small transverse size are expected to
 interact with the nuclear medium
in an anomalously weak manner, due to the above interference
effects~\cite{gribov}.
One typically posits that
any departure from the Glauber estimate is a signal of
color transparency, though other corrections,
such as the inclusion of nuclear correlations~\cite{benhar}, as well as any
change in the assumed reaction mechanism with energy~\cite{svg,bzk},
can modify the nuclear
dependence as well and mimic the excursions arising from
the nucleon's internal structure. Thus, an accurate ``baseline''
calculation in the absence of color transparency effects is essential to
detecting the onset of the phenomenon.

I focus on charmonium production, yet
the issues here --- that is,
the existence and magnitude of any anomalous
nuclear attenuation --- are generic to the diffractive photoproduction of
all vector mesons. The nucleon's internal structure is modelled
in a nonrelativistic constituent quark model, and the computation
performed in the charmonium rest frame, for tractability.
For sufficiently large
quark mass $m_q$,
the energy splitting of the $1S$ and $2S$ quarkonium states
will be small relative to the ground state mass, thus justifying the
nonrelativistic treatment of charmonium's structure.
Thus, the consideration of large quark masses is
convenient theoretically, though
this begs the question of whether a completely consistent calculation
can exist in the large $m_q$ limit, as
the longitudinal momentum transfer to produce other charmonium
states must be able to vanish as the photon energy increases.
These limits are compatible, as
the production and propagation of charmonium in a nucleus
is a two-scale problem.
Crudely, the length
scale for charmonium production is inversely proportional to
the off-shellness of the real photon if it were to fluctuate to a
$J/\psi$, so that the ``creation length'' $l_{\rm c}$
is
\begin{equation}
l_{\rm c} \sim {1 \over {\sqrt{E_{\gamma}^2 + m_{J/\psi}^2} - E_{\gamma}}}
\;,
\end{equation}
whereas the ``formation length'' $l_{\rm f}$, or the inverse
of the longitudinal momentum transfer to produce the next low-lying
charmonium
state,
is given by
\begin{equation}
l_{\rm f} \sim {2p_{J/\psi} \over {(m_{\psi'}^2 - m_{J/\psi}^2) }} \;.
\end{equation}
The  denominator in the above saturates as $m_q$ grows large ---
note that the splitting of the lowest lying $1^{--}$ states is
660 MeV, 588 MeV, and 563 MeV in the $s\bar s$, $c \bar c$, and $b \bar b$
sectors
 --- so that
$l_{\rm f}$ can exceed $l_{\rm c}$  in the large $m_q$ limit.

Here I delineate an
archetypal color transparency calculation~\cite{everyone},
and discuss the various
model estimates and their deficiencies, considering the special needs
of a near-threshold treatment. Certain constraints on the
diffractive charmonium-nucleus
interaction exist as $s\rightarrow\infty$, and
I consider how these constraints may
relax at modest $s$ and what experiments
are germane to elucidating these departures.

\section{An archetypal color transparency calculation}

The $A$-dependence of the nuclear transparency for the production of
charmonium
state $j$, defined as
\be
T_A^j(E_{\rm LAB})
={ \int d^3 r \rho_A(\vec{r})
| \langle \psi_j| \int_z^{\infty} dz' \hat{\sl U}(z',{\vec b})
| \psi_i \rangle |^2
\over
A | \langle \psi_j | \psi_i \rangle |^2 } \;,
\ee
should weaken as the photon energy increases, if color transparency
exists. This suggestion itself assumes
the production mechanism is not intrinsically
$A$-dependent, or, less stringently, that its $A$-dependence is
not energy-dependent~\cite{ralston}.
The above formula is appropriate to charmonium production which is
incoherent from the nucleus, yet coherent from the nucleon, so that
the transverse momentum component of the produced charmonium must be
nonzero, yet less than the pion mass.
The sums over the possible intermediate
nuclear states which exist for every application of $\hat{U}$ are
suppressed --- the momentum of every produced charmonium state is presumed
much larger than the variation in energy of the nuclear intermediate states,
so that the closure approximation can be employed.
Note that $\rho_A (\vec{r})$ denotes the matter density of
the nucleus $A$, $|\psi_i\rangle$ denotes the
initial charmonium
state, $|\psi_f\rangle$ denotes the final charmonium state detected,
and $\hat{\sl U}$ denotes the evolution operator.
Note that the
calculation is realized in the charmonium rest frame.

Certain assumptions underlying Eq.(3) are crudely violated.
Noting Eq.(1), we see that the
reaction mechanism certainly changes with photon energy; the impact
of this change on $T_A$ has been estimated in Ref.~\cite{svg} and
studied in detail in Ref.~\cite{bzk}.
The above formula also models the nucleus as a continuous distribution of
nuclear matter; the inclusion of nuclear correlations can modify the
energy dependence of $T_A$ substantially~\cite{benhar}.
How are the pieces in $T_A$ modelled?

\subsection{$|\psi_f\rangle$: the charmonium state detected.}

The nonrelativistic quark model is
assumed.
Harmonic oscillator potentials have been used
exclusively~\cite{bzk_zak,svg,ben95}: the fits to the charmonium spectrum
are not quantitative.
Consider the $s$-wave charmonium states in the nonrelativistic
quark model. In MeV, we have
\be
\begin{array}{ccccc}
& {\rm EXPT} & {\rm HO}_1 & {\rm HO}_2 & {\rm Richardson} \\
1{\rm S} & 3097 & 3097 & 3097 & 3095 \\
2{\rm S} & 3686 & 3686 & 3569 & 3684 \\
3{\rm S} & 4040 & 4275 & 4040 & 4096 \\
4{\rm S} & 4415 & 4864 & 4512 & 4440
\end{array}
\ee
where ${\rm HO}_1$ denotes a harmonic oscillator fit in which
$2\hbar \omega = m_{\psi(3685)} - m_{\psi(3097)}$~\cite{bzk_zak},
${\rm HO}_2$ denotes the same sort of fit with
$2\hbar \omega = m_{\psi(4040)} - m_{\psi(3097)}$~\cite{svg},
and ``Richardson''
denotes a fit with the Richardson potential~\cite{rich}. The
${\rm HO}_2$ fit is a better caricature of the higher-lying portion
of the charmonium spectrum, though a realistic potential,
such as the Richardson potential,
is preferred. The impact of this simplification could be numerically
large.

\subsection{$|\psi_i\rangle$: the ``initial state''.}

The ``initial state'' is the charmonium production amplitude. Two
different sorts of estimates exist. One estimate uses the amplitude
for the photon to fluctuate to a free $c\bar c$ pair in the infinite
momentum frame times the cross section associated with the
diffractive charmonium-nucleon interaction~\cite{nn_zak}.
This procedure neglects mutual $c \bar c$ interactions, which are equally
important. If they were included in any realistic model, the dependence
on the longitudinal and transverse components of the $c\bar{c}$
separation would not factorize, so that the transformation
to the rest frame, required for the subsequent $T_A$ calculation,
would be unclear. Moreover, the above estimate is
only appropriate in the large photon energy limit.
Other work simply makes an {\it Ansatz} for the initial
amplitude~\cite{bzk_zak,svg} in the near threshold regime;
this is unsatisfying.

Perhaps a better estimate of the production amplitude can
be constructed. Borrowing from threshold
$e^+ e^- \rightarrow t \bar{t}$ studies~\cite{fadin}, the
non-relativistic three-point function for charmonium photoproduction
is given by the Green function
\be
\Big[  - {\Delta \over m_c} + V(\vec{r}_{c \bar c}) - E
\Big]G({\vec r}_{c\bar c}; E) =\delta(\vec{r}_{c\bar c})
\ee
where the source is localized at $\vec{r}_{c \bar c}\,'=0$ and
$V(\vec{r}_{c \bar c})$
describes the mutual $c \bar{c}$ interactions, as well
as the soft interactions with the medium. In the context of top
quark production, this is really a perturbative QCD estimate, as the
enormous decay width associated with the heavy top quark
effectively screens the confining
portion of the potential~\cite{fadin}; here it is a model calculation.

\subsection{${\hat U}$: the evolution operator.}

The quantum mechanical evolution operator is associated with
\be
H(t) = H_{c \bar c} + H_{\rm soft} \;,
\ee
where the time $t$ in the $c \bar c$ rest frame is connected to the
lab coordinate $z$ via $z= v \gamma t$, noting $v$ is the
charmonium velocity and $\gamma$ is the Lorentz factor.
$H_{\rm soft}$ is typically
modelled in a two-gluon exchange picture, so that
the cross section for the $c \bar c$-nucleon
interaction is~\cite{gun_sop,bzk_zak}
\be
\sigma(\rho) = {16 \alpha_s^2 \over 3} \int d^2 {\vec k}
{
[ 1 - \exp(i\vec{k}\cdot \vec{\rho}) ]
[ 1 - F(\vec{k}) ]
\over (k^2 + m_g^2)^2 } \;,
\ee
where $m_g$ is the effective gluon mass, required to regulate
$\sigma(\rho)$ for arbitrary $\rho$, and
$F(k)=\langle N | \exp(i\vec{k}\cdot(\vec{r_1} - \vec{r_2})) | N\rangle$,
where $| N \rangle$ is the nucleon wave function. Finally,
$H_{\rm soft}$ in the charmonium rest frame is chosen to be~\cite{bzk_zak}
\be
H_{\rm soft} = -i {v \gamma \over 2} \sigma(\rho) \rho_A(\vec{r}(t)) \;,
\ee
as the two-gluon exchange amplitude is purely imaginary in the
$s\rightarrow\infty$, $t$ fixed limit. Eq.(8) is not derivable from
Eq.(7); one assumes as per Ref.~\cite{miett} that the bulk of
diffractive charmonium interactions are with the lowest momentum
partons
in the nucleon --- the ``wees'' ---
which are uncontracted even though the nucleon momentum
is large in the charmonium rest frame.
Practically, $H_{\rm soft} \sim C \rho^2$ is chosen, though
corrections have been estimated perturbatively~\cite{ben95}.
There is no reason, however, why the construction of the soft
interaction should stop at two-gluon exchange. Consider three-gluon
exchange, for example. These exchanges are $C$-odd, as well
as $C$-even, in the $t$-channel. The empirical support for the
Pomeranchuk theorem, that is,
$\sigma_{\rm hp}^{\rm tot} \sim \sigma_{\rm \bar{h}p}^{\rm tot}$ as
$s\rightarrow\infty$, can be interpreted to mean that $C$-odd
exchanges are suppressed in the large $s$ limit. Yet the merely
asymptotic approach in $s$ to the Pomeranchuk prediction
indicates in itself that the predictions of the infinite $s$ limit are
modified at finite energies. This is important to the ELFE
energy regime! The consequence is that one has diffractive
production of both natural ($J/\psi, \psi'$) and unnatural ($\chi$)
parity charmonium states in the nucleus, as the final-state nucleons
are undetected. Ref.~\cite{svg} estimates the numerical impact of
the inclusion of both $C$-odd and $C$-even exchanges in the near
threshold regime in an Abelian
string model, though the model is schematic.
In the absence of data, it is difficult to constrain the inclusion of
three-gluon exchange effects.

The $T_A$ calculation can now be effected with the above model
ingredients. The hard part is retaining all the intermediate
charmonium states throughout the calculation.
For harmonic oscillator potentials, one can use the {\it exact}
path integral result for the evolution operator, and the problem
becomes greatly simplified~\cite{bzk_zak}. For other potentials, one
must construct the evolution operator numerically and sum over all
the states~\cite{svg}. Arbitrary two-channel truncations have also
been performed, though they are of unclear reliablity.

The model calculations all show an increase in the nuclear transparency
with increasing photon energy, though a complete calculation in the
near threshold regime has not been performed. On the theoretical
side, a calculation with a quantitative
$H_{c\bar c}$, a production amplitude
estimate as per $G(\vec{r}_{c\bar c}; E)$, and a diffractive interaction
which includes two- and three-gluon exchange is required for an
improved estimate of the nuclear dependence of diffractive charmonium
production in the ELFE energy regime.
On the experimental side, one would like to study the breaking of
the Gribov-Morrison rule, which presumes all diffractively produced
states to have the quantum numbers of the photon, in the nucleus
at moderate to large $E_{\gamma}$.

I thank S. Bass for inviting me to the ELFE summer school and for partial
financial support. This work was supported by the DOE under
Contract No. DE-FG02-87ER40365.


\begin{thebibliography}{10}

\bibitem{brodmuel88} S. J. Brodsky and A. H. Mueller,
Phys. Lett. {\bf B206} (1988) 685.

\bibitem{goodwalk} E. L. Feinberg and
I. J. Pomeranchuk, Suppl. Nu. Cim {\bf
3} (1956) 652; M. L. Good and
W. D. Walker, Phys Rev. {\bf 120} (1960) 1857.

\bibitem{ctbrodmuel} S. J. Brodsky, in {\it Proceedings of the Thirteenth
International Symposium on Multiparticle Dynamic}, (Vollendam, 1982) ed. by
W. Kittel {\it et al} (World Scientific, 1982);
A. H. Mueller, in {\it Proceedings of the Seventh Rencontres
de Moriond} (Les Arcs, France (1982)) ed.
by J. Tr\^an Thanh V\^an (Editions Fronti\`eres,
Gif-sur Yvette 1982).

\bibitem{gribov} V. N. Gribov, Sov. Phys. JETP {\bf 29} (1969) 483.

\bibitem{benhar} O. Benhar,
A. Fabrocini, S. Fantoni, V. R. Pandharipande, and
I. Sick, Phys. Rev. Lett. {\bf 69}, 881 (1992); Nucl. Phys. A {\bf 532}
(1991) 277;
N. N. Nikolaev, A. Szczurek, J. Speth, J. Wambach, B. G.
Zakharov, and V. R. Zoller, Phys. Lett. B {\bf 317} (1993) 287.

\bibitem{svg} S. Gardner, Phys. Rev. C {\bf 48} (1993) 3011.

\bibitem{bzk} B.Z. Kopeliovich, presentation at this workshop.

\bibitem{everyone} P. Jain, J. Ralston, and B. Pire, ``Quantum Color
Transparency and Nuclear Filtering,'' submitted to Phys. Rep.,
November, 1995, and references therein.

\bibitem{ralston} P. Jain and J. P. Ralston,
Phys. Rev. D {\bf 48} (1993) 1104.

\bibitem{bzk_zak} B. Z. Kopeliovich and
B. G. Zakharov, Phys. Rev. D {\bf 44} (1991) 3466.

\bibitem{ben95} O. Benhar, B. G. Zakharov, N. N. Nikolaev, and
S. Fantoni,  Phys. Rev. Lett. {\bf 74} (1995) 3565.

\bibitem{rich} J. L. Richardson, Phys. Lett. {\bf B82} (1979) 272.

\bibitem{nn_zak} N. N. Nikolaev and
B. G. Zakharov, Z. Phys. {\bf C49} (1991) 607;
J. D. Bjorken, J. B. Kogut, and D. E. Soper, Phys. Rev.
{\bf D3} (1970) 1382.

\bibitem{fadin} V. S. Fadin and V. A. Khoze, Sov. J. Nucl. Phys.
{\bf 48} (1988) 309.

\bibitem{gun_sop} J. F. Gunion and D. E. Soper,
Phys. Rev. D {\bf 15} (1977) 2617 and references therein.

\bibitem{miett} H. I Miettinen and J. Pumplin, Phys. Rev. D {\bf 18}
(1978) 1696.


\end{thebibliography}
\end{document}